\documentclass[12pt]{article}
\usepackage[english]{babel}
\usepackage[utf8]{inputenc}
\usepackage[colorinlistoftodos, color=green!40, prependcaption]{todonotes}
\usepackage[pdftex, pdftitle={Article}, pdfauthor={Author}]{hyperref}
\usepackage{subfigure}
\usepackage{authblk}
\usepackage[normalem]{ulem}
\usepackage[absolute]{textpos}
\usepackage{amsmath} % AMS Math Package
\usepackage{amsthm} % Theorem Formatting
\usepackage{amssymb}	% Math symbols such as \mathbb
\usepackage{commath}
\usepackage{mathabx}
\usepackage{lmodern}
\usepackage{fancyhdr}
\usepackage{nameref}
\usepackage{afterpage}
\usepackage{geometry}
\usepackage{fancyhdr}
\usepackage{afterpage}
\usepackage[export]{adjustbox}
\usepackage{wrapfig}
\usepackage[nottoc,numbib]{tocbibind}
\usepackage{afterpage}
\usepackage{braket}
\usepackage{siunitx}
\usepackage{xcolor}
\usepackage{footnote}

\usepackage[backend = bibtex,
      sorting = none, 
      style=numeric-comp,
      url=false,
      isbn=false]{biblatex}
\bibliography{Temp_Inhom_paper}

\newcommand{\microns}[1]{\SI{#1}{\micro\meter}}
\newcommand{\nm}[1]{\SI{#1}{\nano\meter}}

\newcommand{\refeq}[1]{(\ref{#1})}
\newcommand{\refig}[1]{Figure \ref{#1}}

\title{Improved non-linear devices for quantum applications}

\author[1,*]{Jano Gil-Lopez}
\author[1]{Matteo Santandrea}
\author[1]{Benjamin Brecht}
\author[2]{Gana\"el Roland}
\author[1]{Raimund Ricken}
\author[1]{Viktor Quiring}
\author[1]{Christine Silberhorn}
\affil[1]{Integrated Quantum Optics Group, Institute for Photonic Systems (PhoQS), Paderborn University, Warburger Str. 100, 33098 Paderborn, Germany}
\affil[2]{Laboratoire Kastler Brossel, Sorbonne Universit\'{e}, CNRS, ENS-PSL Research University, Coll\`{e}ge deFrance, 4 place Jussieu, F-75252 Paris, France}
\affil[*]{jangil@upb.de}
\date{}

\begin{document}
\maketitle

\begin{abstract}
In this paper, we review the state of the art of mode selective, integrated sum-frequency generation devices tailored for quantum optical technologies. 
We explore benchmarks to asses their performance and discuss the current limitations of these devices, outlining possible strategies to overcome them. Finally, we present the fabrication of a new, improved device and its characterization.
We analyse the fabrication quality of this device and discuss the next steps towards improved non-linear devices for quantum applications.
\end{abstract}

\maketitle

\section{Introduction}

In the last two decades, there has been a significant push towards the implementation of the Quantum Internet \cite{Kimble2008} and quantum computation. 
Different building blocks of these quantum technologies have been investigated: quantum memories and repeaters \cite{Simon2010, Slussarenko2019}, single-photon sources \cite{Eisaman2011}, quantum gates and interfaces \cite{Ansari2018}. 
One of the most studied systems to interface all these components are photons \cite{Zoller2005}: they can be operated at room temperature without suffering from decoherence, can be transmitted through standard optical fiber networks with minimal losses and offer many degrees of freedom to encode information, e.g. polarisation, frequency or phase. 

When choosing the encoding scheme, a high-dimensional one is preferrable, as it offers many advantages, e.g. higher security in quantum key distribution and higher information rates \cite{Cerf2002, Brougham2012, Etcheverry2013, Nunn2013}.
One of the most robust schemes to encode high-dimensional quantum information is that of temporal modes, as they are robust against dispersion in the fiber and naturally provide a high-dimensional basis set.
%These temporal modes can be any set of orthogonal functions spanning the desired Hilbert space and a common choice is the Hermite-Gaussian (HG) functions. 
In this scheme, information is encoded in the temporal degree of freedom of light at infrared wavelengths and then routed via the fiber network to different devices or users. 
To readout quantum information in these temporal modes, a quantum interface that can separately address each temporal mode of the input signal, i.e. that is characterised by single-mode operation, is then necessary. 

Recent years have seen the rise of the quantum pulse gate (QPG)  \cite{Brecht2011} as an ideal single-mode interface to manipulate temporal modes of light.
%A device that accomplishes such a task is the quantum pulse gate (QPG) \cite{Brecht2011}.
%, which  is a tailored sum frequency generation (SFG) process that achieves single-mode operation on temporal modes.
Thanks to a reconfigurable, single-mode transfer function, the QPG can select individual temporal modes from the input signal; the selected mode is up-converted via a sum frequency generation (SFG) process to a shorter wavelength and the part of the signal orthogonal to the transfer function is left unconverted. 
In this way, a QPG device naturally fulfills two independent key requirements for a quantum interface: it allows the communication of quantum optical devices operating at different wavelengths and exploits temporal modes for quantum communication, computation and metrology.
The single-mode operation of the QPG has already been successfully employed in many applications \cite{Ansari2018}, e.g. in quantum state tomography \cite{Ansari2017}, spectral bandwidth compression to interface different quantum systems \cite{Allgaier2017b} and in quantum metrology \cite{Donohue2018, Ansari2021}. 
% For a complete review of the QPG applications and, more generally, tailored SFG processes for quantum technologies, we refer the reader to Ansari et al. \cite{Ansari2018}

To further develop these demonstrations towards everyday applications, efficiency and pure single-mode,this includes spatial and temporal, operation are of utmost importance. However, the ultimate limit to the temporal-spectral performance of the devices are the manufacturing and experimental imperfections, which have not been considered in detail so far.
Characterising, understanding and compensating these imperfections is the key to the improvement of these devices \cite{Santandrea2019}. %\cite{SantandreaGeneralanalysis}
% To this aim, careful characterisation of the devices and the effect of the inhomogeneities \cite{} is key to advance towards realistic, small footprint quantum devices of the future.

In this paper, we review the impact of inhomogeneities in already existing QPG and other SFG based devices.
To characterise the performance of these devices we consider three commonly used benchmarks, namely the \textit{process selectivity}, the \textit{conversion efficiency} and the \textit{bandwidth compression}.
We show that, in the presented cases, these benchmarks have been very far from the theoretical estimations, which can be explained in terms of imperfect fabrication or operation of the devices \cite{Santandrea2019}.
To test this hypothesis, we carefully monitor both the fabrication and the operating conditions of our QPG devices. With a set of improved fabrication parameters and by developing an algorithm to estimate the impact of inhomogoeneities in our systems, we are able to realize a new QPG device with outstanding performance, outperforming former experiments and setting a new frontier for non-linear quantum devices.

\section{Quantum pulse gate benchmarks \label{sec:benchmarks}}

The QPG is an engineered type II SFG process, implemented in titanium in-diffused lithium niobate  waveguides. 
This platform is characterised by a high nonlinearity, allowing extremely efficient frequency conversion processes \cite{Arizmendi2004}, as well as the unique dispersion properties that will be discussed in the following.
%The choice of this platform is motivated by its high non-linear coefficient that provides highly efficient wavelength conversion processes \cite{Arizmendi2004} and the unique dispersion properties, as we will discuss below.
%The process is tailored process is tailored with a periodic poling of \um{4.4}, such that the input (\nm{1550}) and pump (\nm{850}) fields are group velocity matched (GVM) and result in an output (idler) field in the visible (\nm{550}), at an operating temperaure of \SI{200}{^\circCelsius}. The high temperature operation avoids photo-refraction effects allowing high pump field powers for increased efficiency \cite{Augstov1980a}.
In order to achieve temporal single-mode operation, the SFG process is tailored such that the input (\nm{1550}) and pump (\nm{850}) fields are group velocity matched and result in an output (idler) field in the visible (\nm{550}), at an operating temperaure of 200 $^\circ$C and with a poling period of \SI{4.4}{\micro\meter}. The high temperature operation avoids effects arising due to photo-refraction, allowing high pump field powers for increased efficiency \cite{Augstov1980a}.

\begin{figure}[ht]
	\centering
	\includegraphics[width=\linewidth]{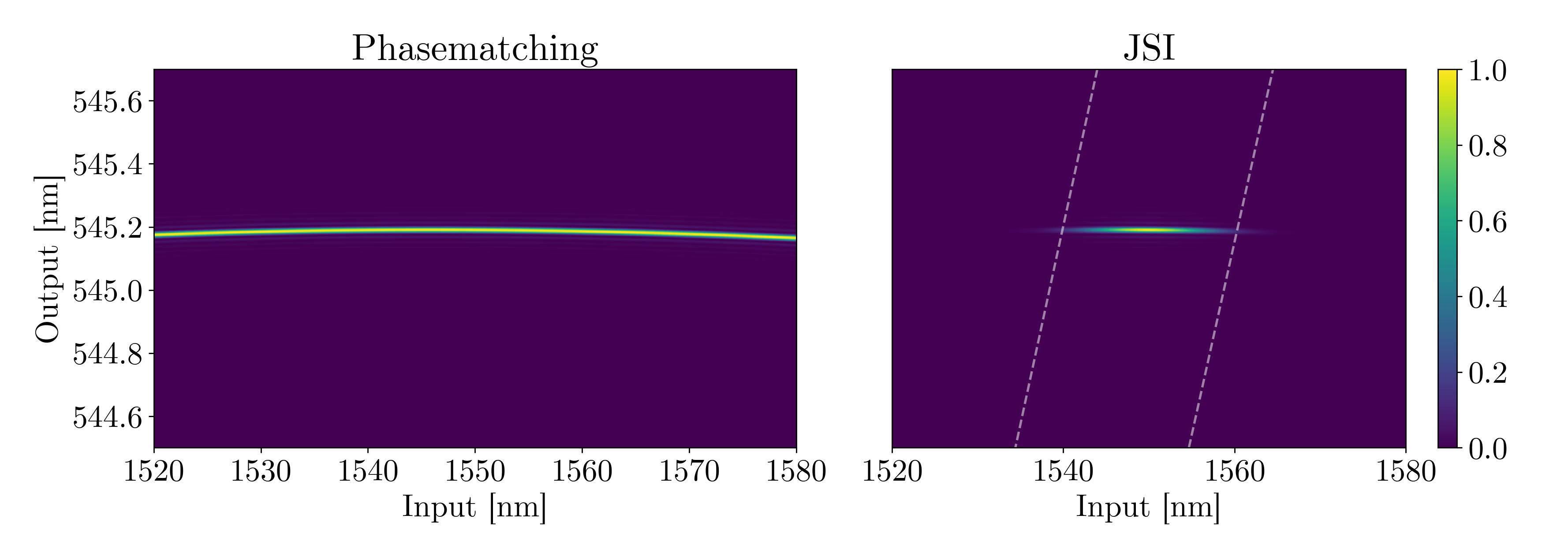}
	\caption{Phase-matching function and JSI of a QPG at 200 $^\circ$C. The pump field used to obtain the JSI is a gaussian pulse centered around \nm{841} with \nm{2.12} $\sigma$ bandiwdth, in dashed lines on the plot.}
	\label{fig:qpg_pm_jsi}
\end{figure}

The group velocity matching (GVM) of input and pump fields is necessary to achieve a flat %phase-matching function and 
joint spectral amplitude/intensity (JSA, JSI), which corresponds to the transfer function of the systems. The flat JSA, depicted in \refig{fig:qpg_pm_jsi}, grants the process single-mode operation and high bandwidth compression \cite{Brecht2011}.

The JSA is the product of the pump field and the phase-matching function

\begin{equation}
    JSA = \phi(\omega_o, \omega_s) \: \alpha(\omega_p), \label{jsa}
\end{equation}

where $\alpha(\omega_p)$ is the pump field envelope and $\phi(\omega_i, \omega_s)$ the phase-matching function, functions of the frequency of the output $\omega_o$, signal $\omega_s$ and pump fields $\omega_p$.

The phase-matching function $\phi$ is completely defined by the wave vector mismatch $\Delta\beta=k_s+k_p-k_o$ of the three fields involved in the SFG process, where $k_s$, $k_p$ and $k_i$ are the wavenumbers of the signal, pump and output fields respectively.  The phase-matching function can be described with \cite{Boyd2008}:

\begin{equation}
	\phi \: \propto \frac{1}{L} \int_{0}^{L} e^{i\Delta\beta(\omega_i, \omega_s, \omega_p) z} dz \rightarrow \mathrm{sinc}\left( \frac{\Delta\beta(\omega_o, \omega_s, \omega_p) L}{2} e^{i\frac{\Delta\beta(\omega_o, \omega_s, \omega_p) L}{2}} \right), \label{ideal_pm}
\end{equation}
where L is the length of the waveguide and z the propagation axis.

\begin{figure}[ht]
	\centering
	\includegraphics[scale=1]{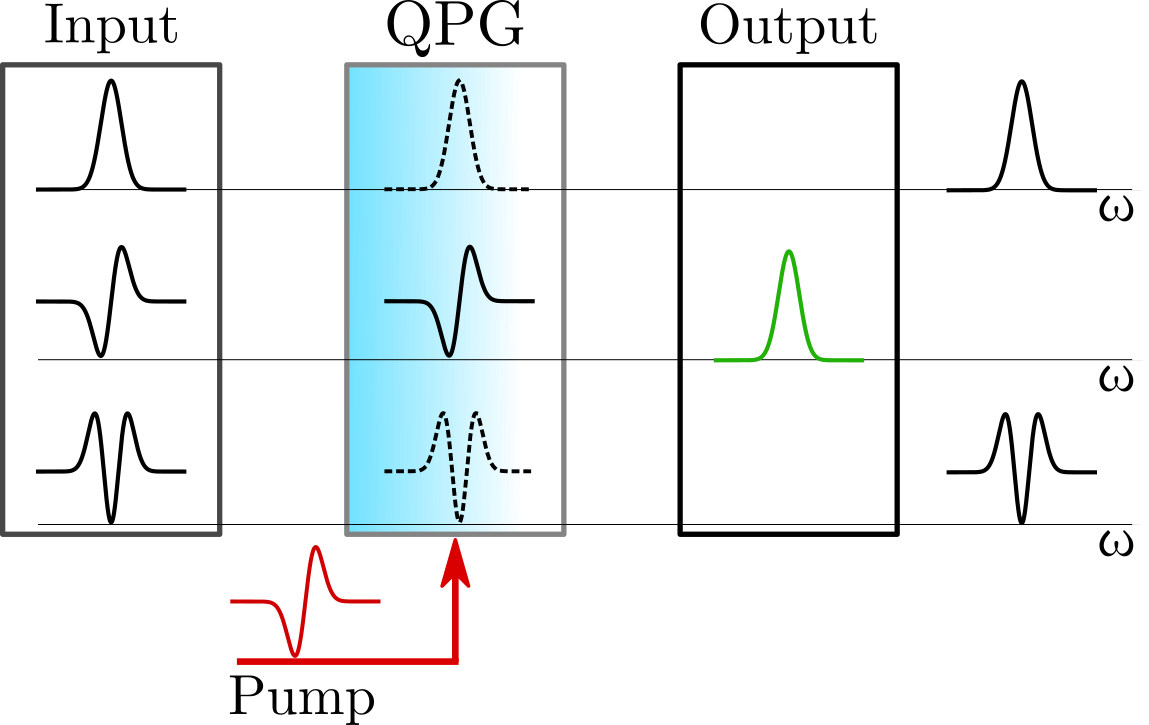}
	\caption{The QPG performs SFG on the input modes that overlap with the pump modes, the remaining modes propagate unperturbed.}
	\label{fig:qpg_sketch}
\end{figure}

% The flat phase-matching function, that results from the unique dispersion properties of lithium niobate, allows for a single-mode operation on temporal modes of the input and output fields. 
The single-mode transfer function of the QPG can be easily reconfigured by spectral shaping of the pump field $\alpha(\omega_p)$ into different temporal modes. 
In particular, only the components of the input signal that overlap in spectrum and time with the chosen pump modal distribution are converted to the output.
The mode selective process and a examples of temporal mode selectivity using the  Hemite-Gaussian basis set are sketched in \refig{fig:qpg_sketch}.

From this discussion, it is clear that single-mode operation depends on the phase-matching function.
Therefore, achieving the ideal phase-matching function is necessary to push the device performance to its limit. 
Unfortunately, direct comparison of the phase-matching spectra of different devices is not trivial, as it is hard to find a single metric to quantify the spectral differences between different samples. Therefore, we will consider in the following three common performance metrics used in the field, namely the \textit{conversion efficiency}, the \textit{bandwidth compression} and the \textit{temporal mode selectivity}. Their relation to the phase-matching properties is discussed in the following paragraphs.

\subsection*{Conversion efficiency}

The conversion efficiency is measured by taking the ratio of output, converted power to input signal power.
It can be theoretically estimated by solving the coupled mode equations of the three fields interacting through the waveguide in the non-linear material.
Assuming no pump depletion, the efficiency $\eta$ of a SFG process in a waveguide with length L and excited with pump power $P_p$ scales as \cite{Boyd2008}

\begin{equation}
    \eta = \frac{P_{converted}}{P_{input}} = \sin^2\sqrt{\eta_{norm} P_p} L \label{eta},
\end{equation}

with $\eta_{norm}$ the efficiency of the process, normalized per unit pump power and sample length \cite{Cerullo2003}. 
The normalized efficiency depends on the effective non-linear coefficient for the process considered, the spatial overlap, the effective refractive indices and frequencies of the interacting fields.
For ultrafast pulses, it also depends on the characteristics of the pulses.
Therefore, under the same experimental conditions, i.e. identical nonlinear crystal and pulse properties, the only way to increase the conversion efficiency of the QPG is by increasing its length. However, as we will discuss later on, this is not always straightforward, as longer samples are more sensitive to fabrication imperfections and experimental instabilities.

\subsection*{Bandwidth compression}

As previously mentioned, the flat JSA allows the conversion of broadband, input signals in the infrared to output pulses in the green. 
Bandwidth compression is achieved when the bandwidth of the output field is smaller than the input one.
If the QPG process is properly engineered, it results in a more efficient bandwidth compression than could be achieved by simple spectral filtering \cite{Allgaier2017b}.
%If the efficiency of the non-linear process and the losses of the experimental setup are lower than the losses of an equivalent spectral filter, a wide input will be filtered down by the JSI to a narrow output without loss of information and more efficiently than simply using an comparable filter \cite{Allgaier2017b}. 

For a given material, the output bandwidth is determined by the length of the sample. In contrast, the input bandwidth depends on the signal that must be compressed for the communication protocol.
Therefore, assuming a well-defined, constant input signal, the bandwidth compression is directly related to the width of the phase-matching function and can be improved by employing longer samples \cite{Santandrea2019}.
%Therefore, for a fixed input, the bandwidth compression can be increase by designing and optimising longer samples \cite{Santandrea2019}.

% From the JSI we can obtain the expected bandwidth of the output field.
% The bandwidth compression factor is then given by the ratio of the bandwidths of a certain input and of the output on the JSI.
% For the same input signal, the compression ratio will increase with smaller output bandwidth.
% The bandwidth can be reduced by increasing the length of the waveguide, improving the bandwidth compression factor.

\subsection*{Mode selectivity}
The QPG has been described so far as a device implementing an ideal single-mode transfer function.
However, this is only an approximation. The frequency correlations in the  pump spectrum, related to energy conservation, are imprinted on the transfer function of the system and reduce the performance of the QPG.
In particular, they increase the number of modes of the transfer function. 
The net result is that spectral components of the input signal that are orthogonal to the pump spectrum are converted by the non-ideal transfer function of the system, leading to cross-talk between orthogonal temporal modes during the frequency conversion operation.
%HIGH ORDER PUMP --> MORE DIAGONAL FEATURES --> CANNOT BE PROJECTED TO A SINGLE MODE FOR THE INPUT.

%IF THE PM BANDWIDTH IS REDUCED, LESS PROMINENT DIAGONAL FEATURES --> CLOSER TO A SINGLE MODE

The amount of cross-talk can be quantified using the final benchmark, mode selectivity. It quantifies the portion of the input field upconverted from the temporal modes orthogonal to the desired one.
% The ideal QPG device, as described in \refig{fig:qpg_sketch}, yields a mode selectivity of 1. On the JSI, the selectivity is given by the overlap of the input field with the JSI.

Analysing the JSA and its underlying Schmidt mode structure provides information on the temporal mode selectivity. From this analysis, the selectivity can be calculated from \cite{Reddy2013a}

\begin{equation}
    S := \frac{|\rho_m|^2}{\sum_{n=0}^{\infty}|\rho_n|} \label{selectivity},
\end{equation}

where $m$ is the desired mode, $n$ is the order of all the modes in the JSA and $\rho_n$ are the Schmidt coefficients for each  mode, normalised to $\sum_n \rho_n=1$.
The selectivity, as defined in Eq. \eqref{selectivity}, is hard to quantify experimentally. 

On the other hand, the modal extinction ratio, defined as $\varepsilon=-10\: log_{10}(\sqrt{S})$, can be retrieved  experimentally from the up-converted intensities, when the process is pumped with different orthogonal temporal modes and a fixed signal. 
We can measure the output field intensity $P$ produced by the fixed signal on the first temporal mode and changing the pump mode from that of the signal to the next orthogonal mode. 
With each pump mode, we obtain the up-converted intensity $P_0$ and $P_1$, where the subindex is the order of the mode.

The ratio $P_1/P_0$ between the up-converted intensity in both cases is the separability from which we obtain the extinction ratio, used from now on to characterise the mode selectivity of the process. 
Experimentally, it suffices to approximate \refeq{selectivity} with just the first two modes of the Schmidt distribution as higher modes will be hindered by dark counts and noise of the detection system.

In practical applications, one generally cannot achieve single-mode operation.
Since the pump is aligned at +45$^\circ$ (for energy conservation), shaping the pump with higher order HG modes introduces spectral correlations. 
%Therefore, pure single mode operation is impossible in our samples. 
This can be greatly mitigated by narrowing the PM bandwidth.

\vspace{5mm}

%The benchmarks discussed in the previous paragraphs, i.e. the process efficiency, selectivity and bandwidth compression, are all related to the properties of the phase-matching spectrum. 
As we can conclude from our discussion, the properties of the phase-matching function are decisive for all benchmarks, namely the process efficiency, the selectivity and the bandwidth compression. All of them can be can be dramatically improved by reducing the phase-matching bandwidth, i.e. by increasing the waveguide length.

As we will review in the next sections, manufacturing long waveguides has limited results in the past decade. 
In recent years, the effects of waveguide inhomogeneities have been studied in detail and found to be the possible source of many experimental sources of error. 
Santandrea et al. \cite{Santandrea2019, Santandrea2019a} studied the effects of waveguide inhomogeneities in titanium in-diffused lithium niobate waveguides and, in particular, their connection to variation of the $\Delta\beta$ along the waveguide, which is related to the spectral quality of the phase-matching function.
Since longer devices are required to improve the given benchmarks, the impact of the inhomogeneities is higher and a more careful manufacturing and characterisation is needed to achieve the maximum performance.

\section{State of the art and future perspectives}

In the last decade, research groups have found highly efficient devices with good performance that already allow for proof of concept of real applications: in metrology \cite{Donohue2018}, interfacing quantum memories \cite{Allgaier2017b}, manipulation of temporal modes \cite{Ansari2017, Kowligy2014} and new photon detection schemes \cite{VanDevender2003, Pelc2011}. 
In table \ref{tab:state_art}, results from different publications using the QPG process or similar are listed. 

Conversion efficiency has consistently improved over the years, reaching the predicted maximum efficiency of 80\%, limited by time-ordering corrections \cite{Quesada2016}.

Regarding the achieved bandwidths and selectivity, the current limits are \SI{0.1}{\nano\meter} and around \SI{7}{\deci B} respectively. 
In \cite{Donohue2018}, the output bandwidth of the device was tightly filtered down to the reported bandwidth to increase the selectivity. 
However, this severely reduced the external conversion efficiency and renders the process useless for quantum state inputs and bandwidth compression. 

The current high-performance QPG devices have always been shorter than \SI{30}{\milli\meter}. Moreover, their performance is usually quite away from the theoretically expected one.
Nevertheless, the systems presented in Table \ref{tab:state_art} show that these devices work in principle, although further optimisation is necessary to increase their applicability in future quantum technologies. 

\begin{figure}[ht]
	\centering
	\includegraphics[width=\linewidth]{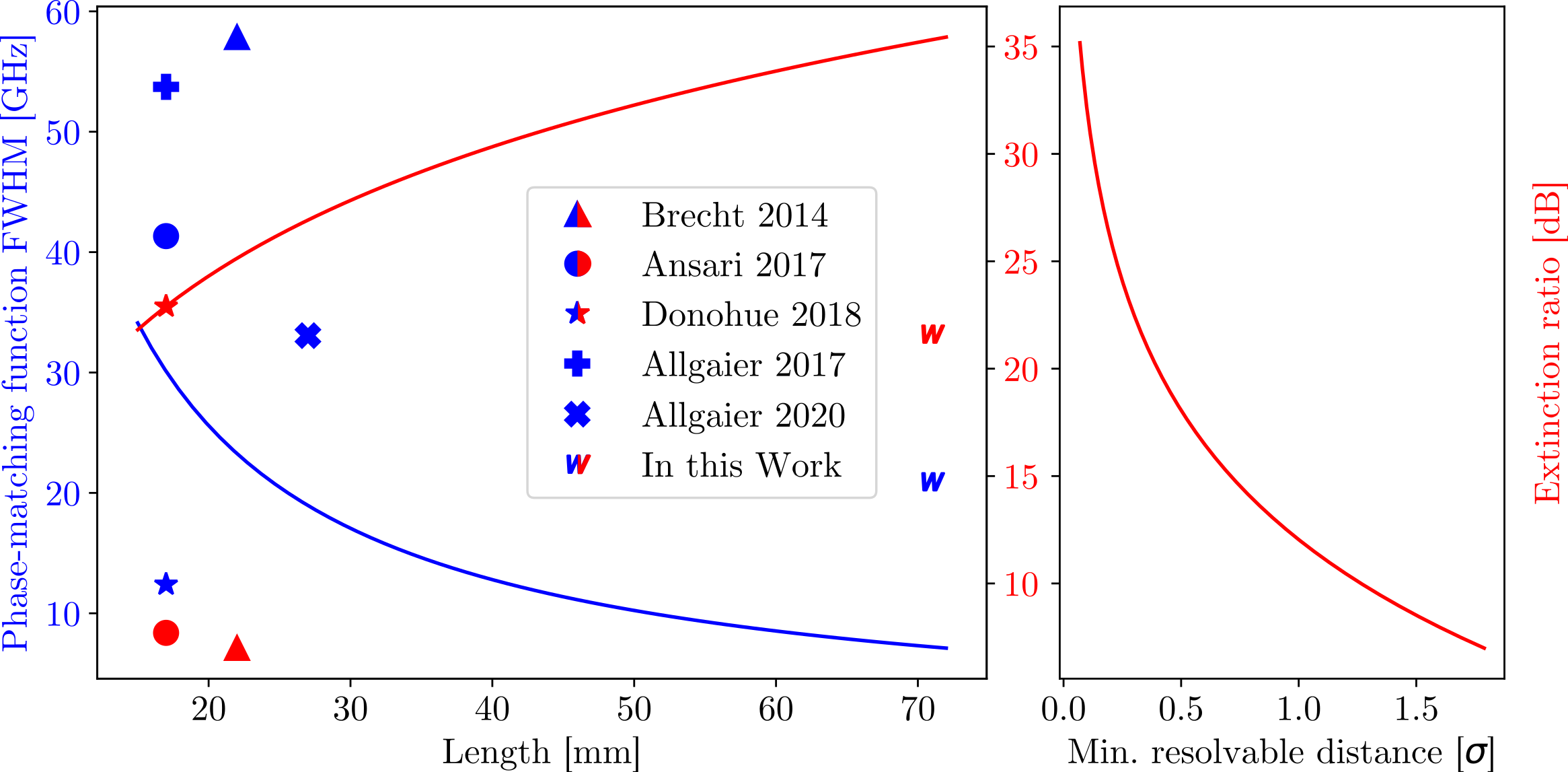}
	\caption{a) Expected phase-matching function full width at half maximum (FWHM) (blue) and extinction ratio (red) for the length of the device. The reviewed articles and the results presented in this paper are shown for comparison. b) Minimum resolvable distance in terms of the $1/e$ bandwidths ($\sigma$) of two close signals expected for the extinction ratio.}
	\label{fig:bw_L_select_minsigma}
\end{figure}

We calculated the expected bandwidth and extinction ratios for the QPG process in lithium niobate waveguides. 
The published results from Table \ref{tab:state_art} are compared to the theoretically achievable values in \refig{fig:bw_L_select_minsigma} a). One immediately sees that device performance is, in general, far from the theoretical limit; the spectral bandwidths are always much wider than expected and the selectivities lower than estimated. The only exception are the results by Donohue et al. \cite{Donohue2018}, where a tight spectral filter was used to improve the performance of the process at the expense of efficiency.
%The results can be seen in \refig{fig:bw_L_select}, compared to the ones reviewed in this section. 
%The first thing that the calculations reveal is that the already existing devices are very far off the expected bandwidth for their length and therefore the extinction ratio is severely reduced too. 
The widening of the bandwidth is usually the result of varying phase-matching condition along the waveguide. This is usually connected to imperfect waveguide fabrication or non-ideal operating condition of the device \cite{Santandrea2019}. 

\begin{savenotes}
    \begin{table}[ht]
        \centering
        \label{tab:state_art}
        \begin{tabular}{l|l|l|l|l|l|l}
        \multicolumn{1}{c|}{\begin{tabular}[c]{@{}c@{}}Length\\{[}mm]\end{tabular}} & \multicolumn{1}{c|}{\begin{tabular}[c]{@{}c@{}}Output\\Bandwidth\\{[}nm]\end{tabular}} & \multicolumn{1}{c|}{\begin{tabular}[c]{@{}c@{}}Selectivity\\{[}dB]\end{tabular}} & \multicolumn{1}{c|}{\begin{tabular}[c]{@{}c@{}}Bandwidth\\compression \end{tabular}} & \multicolumn{1}{c|}{\begin{tabular}[c]{@{}c@{}}Internal\\conversion\\efficiency \end{tabular}} &
        \multicolumn{1}{c|}{\begin{tabular}[c]{@{}c@{}}$\eta_{norm}$\\{[}W$^{-1}$cm$^{-2}$]\end{tabular}} &
        Ref.                \\ 
        \hline
        22                                                                                                             & 0.14                                                                                                                      & 7                                                                                                                   & 6.98                                                                                 & 87\%
        & -
        & \cite{Brecht2014}       \\
        27                                                                                                             & 0.13                                                                                                                      & -                                                                                                                   & 7.47                                                                                 & 61.5\%                                               & 2.32                                         
        & \cite{Allgaier2017b}    \\
        17                                                                                                             & 0.1                                                                                                                       & 7.7                                                                                                                 & 9.71                                                                                 & 5\% 
        & -
        & \cite{Ansari2017}       \\
        17                                                                                                             & 0.03                                                                                                                      & 22.9                                                                                                                & -                                                                                    & 18\%
        & 3.32
        & \cite{Donohue2018}      \\
        27                                                                                                             & 0.08                                                                                                                      & -                                                                                                                   & -                                                                                    & -  
        & -
        & \cite{Allgaier2020}     \\
        15                                                                                                             & -                                                                                                                         & -                                                                                                                   & -                                                                                    & 80\%       
        & -
        & \cite{VanDevender2003}  \\
        47                                                                                                             & -                                                                                                                         & -                                                                                                                   & -                                                                                    & 86\%    
        & -
        & \cite{Pelc2011}         \\
        60                                                                                                             & -                                                                                                                         & 8.4                                                                                                                 & -                                                                                    & 80\%   
        & -
        & \cite{Kowligy2014}    \\
        71                                                                                                             & 0.03                                                                                                                         & 21.5\textsuperscript{*}                                                                                                                 & 16                                                                                    & 18\%\textsuperscript{**}   
        & 1.07
        & -
        \end{tabular}
        \caption{Results from different on non-linear frequency conversion. The bandwidth compression has been calculated with an input of \SI{963}{\giga\hertz}. Normalized conversion efficiency $\eta_{norm}$ has been estimated for the processes where the pulse properties were known. The last entry of the table contains the results presented in this paper. \textsuperscript{*}Estimated from experimental results, see discussion in section \ref{sec:steps}; \textsuperscript{**}at maximum available power, see discussion in section \ref{sec:steps}}.
    \end{table}%
\end{savenotes}

As discussed in \cite{Santandrea2019}, for a given technology and operating conditions, a nonlinear process is characterised by a well-defined \textit{critical length}. Samples longer than this critical length are much more prone to exhibit non-ideal phase-matching spectra and reduced conversion efficiencies. In particular, for the  QPG device under investigation, Santandrea et al. \cite{Santandrea2019} fix the critical length between 1 and 2 cm. To increase this critical length it is usually necessary to dramatically improve the fabrication technology and/or the operating conditions of the device, which is a non-trivial task.
However, the benefits stemming from correcting or compensating these effects justify the efforts toward the continuous improvement of the fabrication and operation of these samples.

If bandwidth compression is improved and the output bandwidth reaches values around a few tens of \SI{}{\giga\hertz}, most Raman quantum memories could be easily interfaced with single photon sources \cite{Allgaier2017b, Heshami2016}. 
Several advantages are to be expected also when using the QPG for metrology applications \cite{Donohue2018, Ansari2021}. 
For example, smaller bandwidths would greatly increase the extinction ratio and therefore the minimum estimable separation between signals when using the QPG as a device to estimate the separation of two very close signals, as shown in the calculations in \refig{fig:bw_L_select_minsigma} b).

% In this metrology scheme, the minimum estimable spectral distance between the two fields depends on the extinction ratio of the QPG. In \refig{fig:min_sigma} the calculated minimum estimable separation between signals is plotted against the extinction ratio.

Since the theoretical estimations presented in \refig{fig:bw_L_select_minsigma} have not been reached yet, it is clear that the current fabrication technology and the operation of these systems can still be greatly improved. 
%  If the quality of the devices is improved or the effects of the inhomogeneities compensated, much narrower bandwidths would be reached. 
 
%  Correcting the errors on the current devices to reach the expected values, every benchmark would be at least twice better. 
%  Furthermore, perfect devices longer than \SI{30}{\milli\meter} would improve the results even further: doubling the length to \SI{60}{\milli\meter} would halve the bandwidth again and increase the extinction ratio to around \SI{40}{\deci B} for a minimum estimated distance smaller than 0.1 sigma. 

\section{Steps toward improved devices \label{sec:steps}}

To increase the length of our samples and achieve the expected bandwidth and extinction ratio, it is essential to identify and reduce the primary sources of phase-matching variation along the waveguide. 
Such inhomogeneities can be caused by errors in the width and depth of the waveguide, purity of the titanium or gradients during the diffusion process, but also uneven temperature distribution of the device when in use.
We start this investigation by focussing on improving the fabrication process.
To this aim, a new photolithography masks with more uniform poling patterns and waveguide structures has been produced in order to minimise poling errors due to uneven current densities during electric field poling; the use of a pristine titanium batch was ensured; the uv-lamp was carefully adjusted for more homogeneous illumination of the photolithography mask. 
Several devices have been produced, where the poling charge has been varied, to test its effect on the poling patters. The best results were found for the sample that had been poled with double the electrical charge required for domain inversion. After these modifications, we manufactured \SI{70}{\milli\meter} long QPG devices and characterised them.

\begin{figure}[h!]
	\centering
	\includegraphics[scale=0.95]{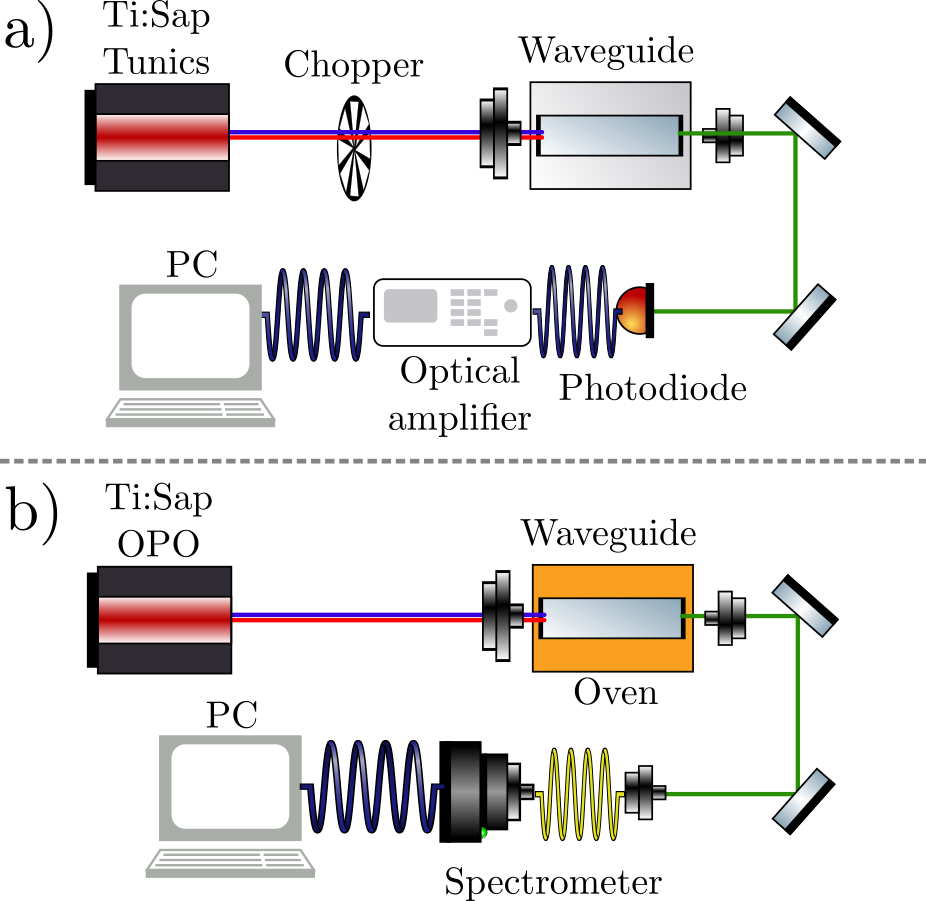}
	\caption{a) Experimental setup to measure the phase-matching of the sample at room temperature. The signal is a tunable Tunics CW laser. The pump source is a Ti:sap CW laser. Both are coupled into the sample and the \nm{550} output intensity is measured with a photodiode. An optical amplifier is used in combination with a chopper to amplify the signal and the data is stored on a computer. b) Experimental setup to measure the phase-matching of the sampple at 200 degrees Celsius. The signal comes from an OPO system and the pump from the Ti:sap that pumps it. This are ultrafast pulses of \SI{200}{\femto\second} duration. The fields are coupled into the waveguide and the generated field is measured on a fiber-coupled spectrometer.}
	\label{fig:exp_setup}
\end{figure}

To quantify the improvements, we characterise a \SI{71}{\milli\meter} long lithium niobate sample with titanium indiffused waveguides, with a poling period of \microns{4.4}. 
The experimental setup used to measure the phase-matching function at room temperature is shown in \refig{fig:exp_setup}.

\begin{figure}[h!]
	\centering
	\includegraphics[width=\linewidth]{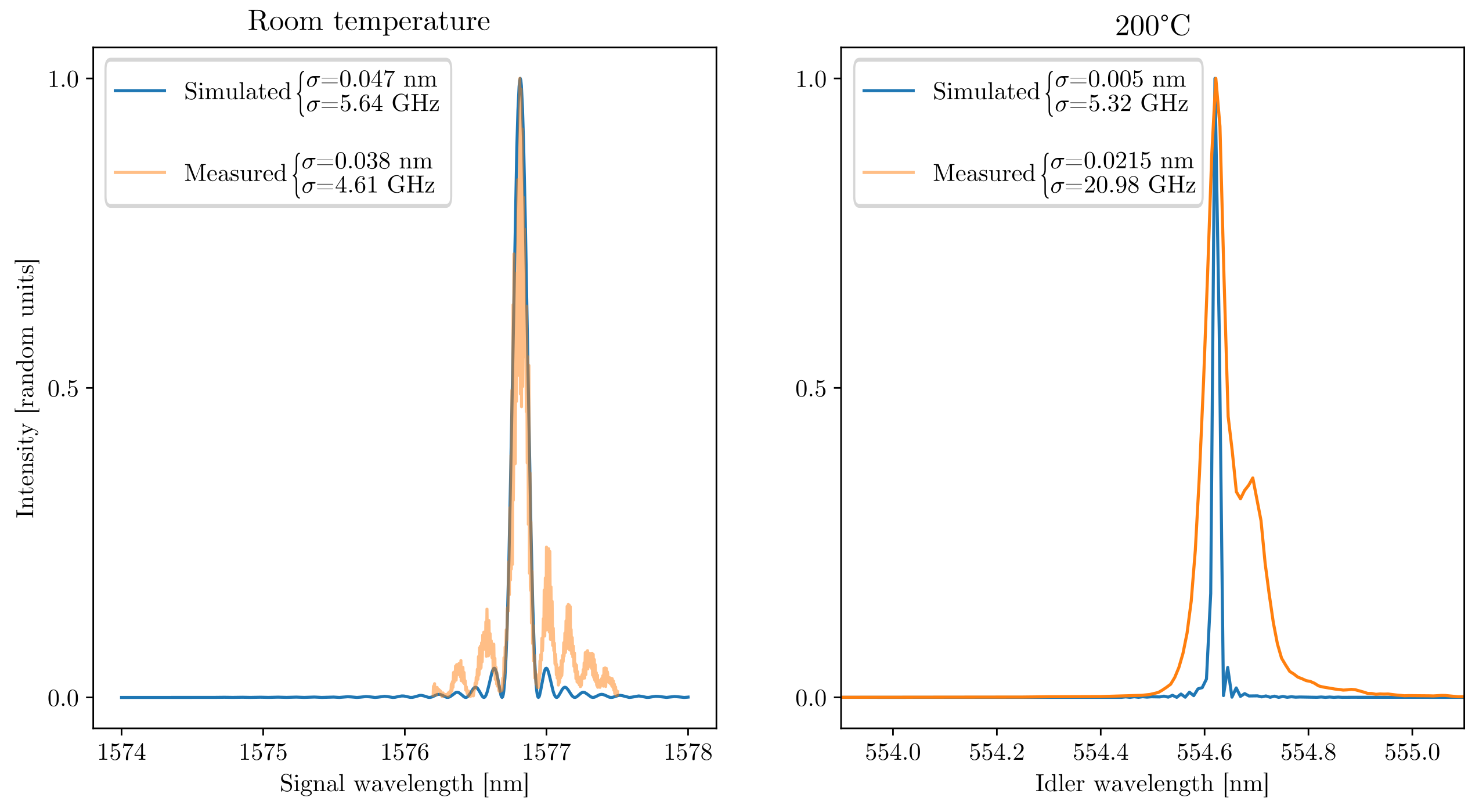}
	\caption{Simulated and measured phase-matchings of the sample at room temperature ($\approx$21$^\circ$C) and at 200 $^\circ$C. At 200 $^\circ$C the simulation was performed with the spectrometer's resolution. The $1/e$ bandwidth ($\sigma$) of each spectra is annotated in the legend in wavelength and frequency units.}
	\label{fig:pm_RT_200}
\end{figure}

The waveguide is pumped with both a signal and a pump field. The signal wavelength is scanned from 1574 to \nm{1578} with a fixed pump at \nm{875}.
The intensity of the generated SFG field is then recorded at each signal wavelength.
The result is shown on the left plot in \refig{fig:pm_RT_200}. 
The measured phase-matching exhibits a bandwidth of $\sigma =$ \nm{0.47}, very close to the expected one. 
The measured spectrum is slightly narrower than the theoretical one due to the presence of the interference fringes caused by the measurement technique.
%The slight difference can be explained by the interference fringes caused by the measurement technique \cite{Santandrea2020}. 
The overall symmetry and side lobes don't fit. This is due to fabrication errors on the waveguide. 
Altogether, this is an outstanding result for such a long waveguide, as the expected critical length for the process, characterised in \cite{Santandrea2019}, is below \SI{20}{\milli\meter}.
    
\begin{figure}[h!]
	\centering
	\includegraphics[scale=0.4]{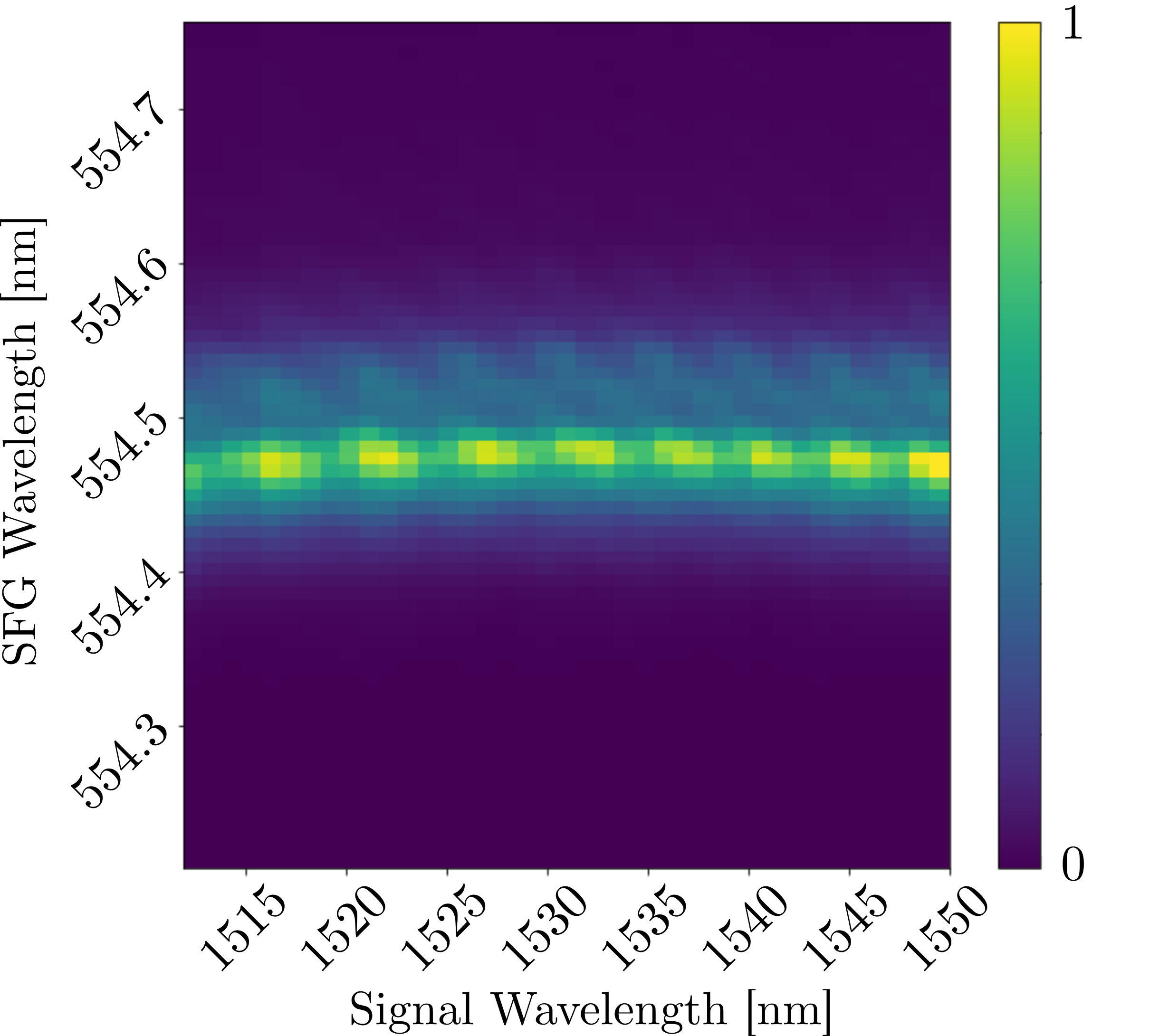}
	\caption{2D phase-matching of the QPG process measured at 200 $^\circ$C. The intensity beating seen is an artifact of the experimental measurement.}
	\label{fig:2d_pm}
\end{figure}

With the sample performing close to expected at room temperature, the phase-matching spectrum is then characterised at 200 $^\circ$C on the experimental setup depicted in \refig{fig:exp_setup}b). 
In this setup, the sample is heated up to 200 $^\circ$C inside a copper oven. 
The oven is kept at constant temperature with a heating cartridge controlled on a feedback loop by a temperature controller (Oxford instruments Mercury iTC). 
Ultrafast pulses at the signal and pump wavelengths, with a duration of \SI{200}{\femto\second}, are launched into the sample and the generated field is directly measured on a spectrometer (Andor Newton EMCCD, 2400 lines per mm grating, resolution ($\approx$\SI{0.03}{\nano\meter})). 
Note that the spectrometer resolution allows us to measure reliably phase-matching spectra from samples around \SI{40}{\milli\meter} long.  %around \nm{0.03}. %of samples up to \SI{40}{\milli\meter}. 
The recorded spectra is presented in a 2D plot in \refig{fig:2d_pm} and a measured 1D slice of it is shown on the right plot on in \refig{fig:pm_RT_200}.

The 1D phase-matching function at 200 $^\circ$C shows a narrow, slightly asymmetric spectrum, close to the expected one. 
The 1/e bandwidth of the measured SFG field, corrected for the resolution of the spectrometer, is $\sigma =$ \nm{0.0215} while the expected value is $\sigma =$ \nm{0.005}. 
The side lobes present at room temperature have disappeared and the main peak displays an asymmetry that will impact the performance of the QPG. 
Nevertheless, the achieved bandwidth is much narrower than any previous results. %and could be smaller than measured due to the resolution of the spectrometer.

\begin{figure}[h!]
	\centering
	\includegraphics[width=\linewidth]{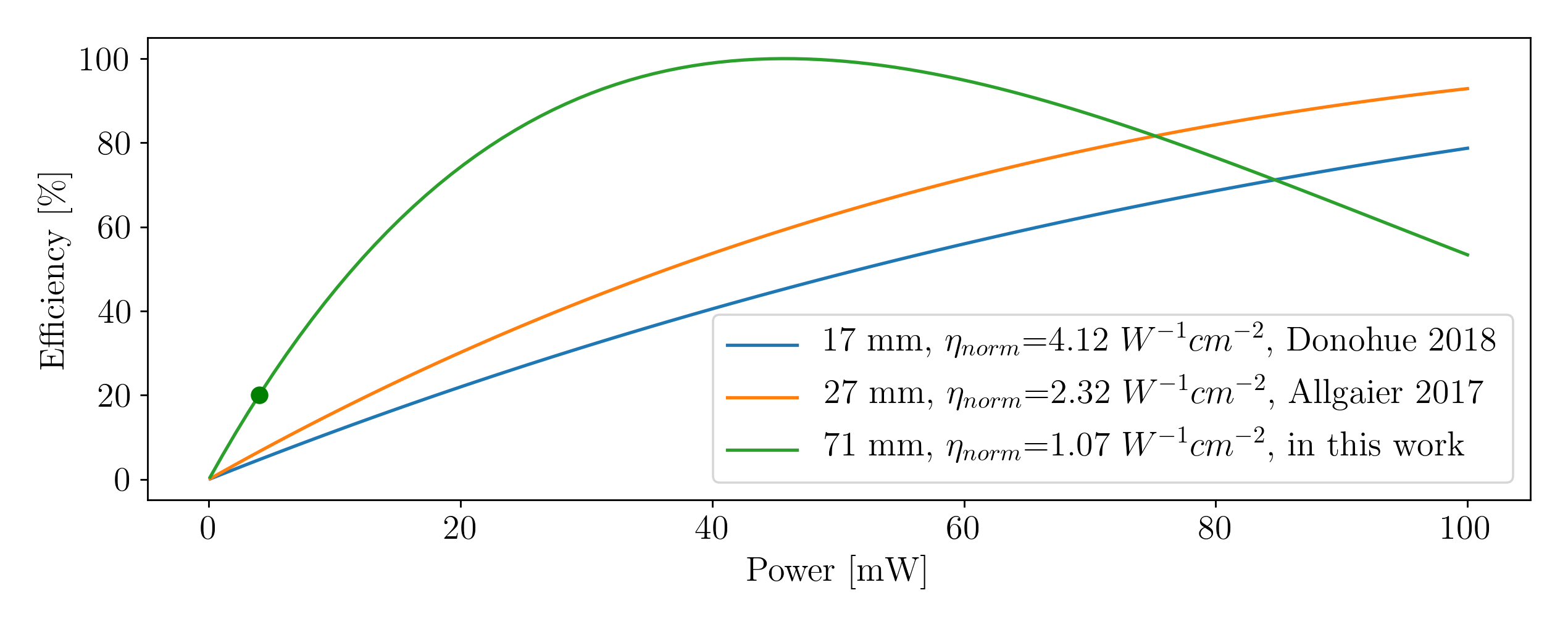}
	\caption{Comparison between the conversion efficiencies calculated from the $\eta_{norm}$ in \cite{Allgaier2017b, Donohue2018} and in this paper with equation \ref{eta}. The green dot marks the value measured experimentally.}
	\label{fig:effcy}
\end{figure}

The internal conversion efficiency was measured by recording the depletion of the signal power through the device for different pump powers. 
With the maximum available power, \SI{4}{\milli\watt}, the conversion efficiency was 18\%.
The normalised conversion efficiency, calculated from \refeq{eta}, is \SI{1.15}{\watt^{-1}\centi\meter^{-2}}. 
With respect to previous results \cite{Allgaier2017b}, we measure much higher internal conversion efficiency, even with lower $\eta_{norm}$.
%This can be explained taking into account that the sample is four times longer.
Therefore, the reduction of $\eta_{norm}$ due to fabrication imperfection is more than compensated by a fourfold increase in the interaction length \cite{Santandrea2019}. 
The enhancement is evident when one calculates the conversion efficiencies from equation \ref{eta} for the $\eta_{norm}$ of the reviewed experiments, the length increase allows to reach unit efficiency at much lower pump powers than previous shorter samples. This can be seen in figure \ref{fig:effcy}.

The phase-matching function at room temperature showed a nearly ideal main peak. However, we observe some imperfections in the phase-matching spectrum at 200 $^\circ$C, namely slightly wider bandwidth and a secondary peak.
A natural question is whether the spectrum at 200 $^\circ$C is consistent with the one at room temperature. 
To this aim, one can estimate the inhomogeneities at room temperature and see if they fit at high temperature. 
To do this, it is necessary to retrieve the $\Delta\beta$ variation along the waveguide at room temperature and use it to simulate what should happen at 200 deg.

% These inhomogeneities, combined with the higher temperature, might the explain the phase-matching function at \SI{200}{^\circCelsius}.

%There exists involved experimental methods to measure the profile of the inhomogeneities on waveguides such as the one described by Chang et al. \cite{Chang2014}. 
Although it is possible to measure the profile of waveguide inhomogeneities, the required experimental setups are generally very complicated \cite{Chang2014}.
This method requires its own dedicated experimental setup and a Frequency Resolved Optical Grating (FROG) system to reconstruct the $\Delta\beta$ profile of the structure.  

A different approach is to obtain the $\Delta\beta$ profile directly from the phase-matching function measurement. 
To do this, we describe the phase-mismatch profile as a function of the propagation axis $z$, dividing the waveguide into m sections with constant $\Delta\beta$ given by $f_m(z)$: $\Delta\beta(f_m(z))$. 
Then, by tuning the $\Delta\beta$ value of each section of the profile, we can find the profile that generates the best fit to the measured phase-matching function.

\begin{figure}[h!]
	\centering
	\includegraphics[scale=0.5]{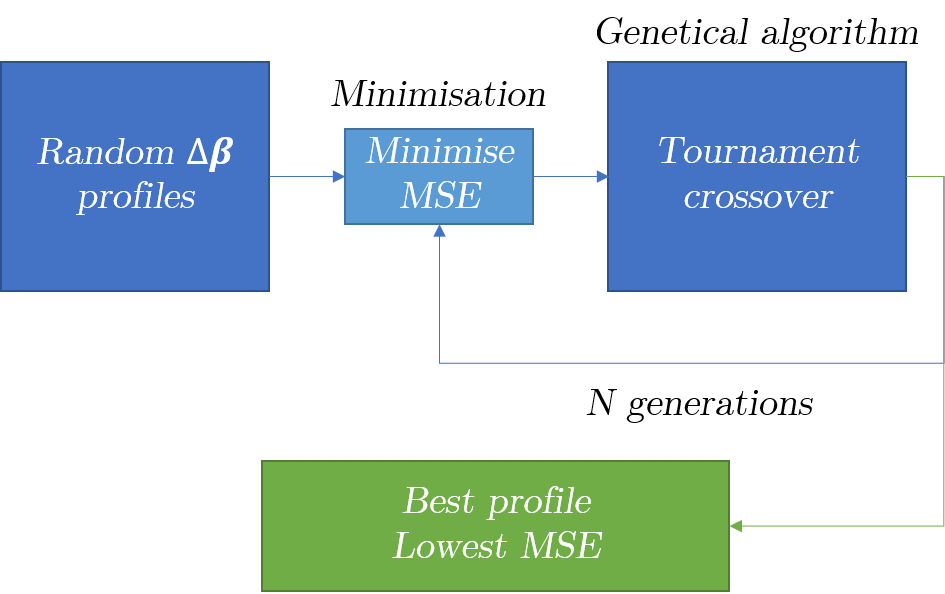}
	\caption{Flowchart of the optimisation algorithm.}
	\label{fig:algorithm}
\end{figure}

To find the best $f_m(z)$ for each section we use a genetic algorithm combined with a minimisation routine. 
We look for the profile that results in a phase-matching function that minimises the mean square error (MSE) between the simulated and the measured room temperature phase-matching spectra.
%We look for the profile that fits the measured phase-matching at room temperature with the minimum mean square error (MSE). 
The algorithm uses the MSE between the phase-matching functions produced by randomly generated $\Delta\beta(f_m(z))$ profiles and the measured phase-matching as a fitting function. 
Each profile's MSE is first minimised using the Broyden-Fletcher-Goldfarb-Shanno algorithm and then some low MSE profiles are selected with a tournament selection rule \cite{Fang2010}. 
The selected profiles are crossed over to generate new profiles. 
This process is repeated for a number of generations to reduce the MSE of the best profile. 
A flowchart describing the algorithm is shown in \refig{fig:algorithm}.

\begin{figure}[h!]
	\centering
	\includegraphics[width=\linewidth]{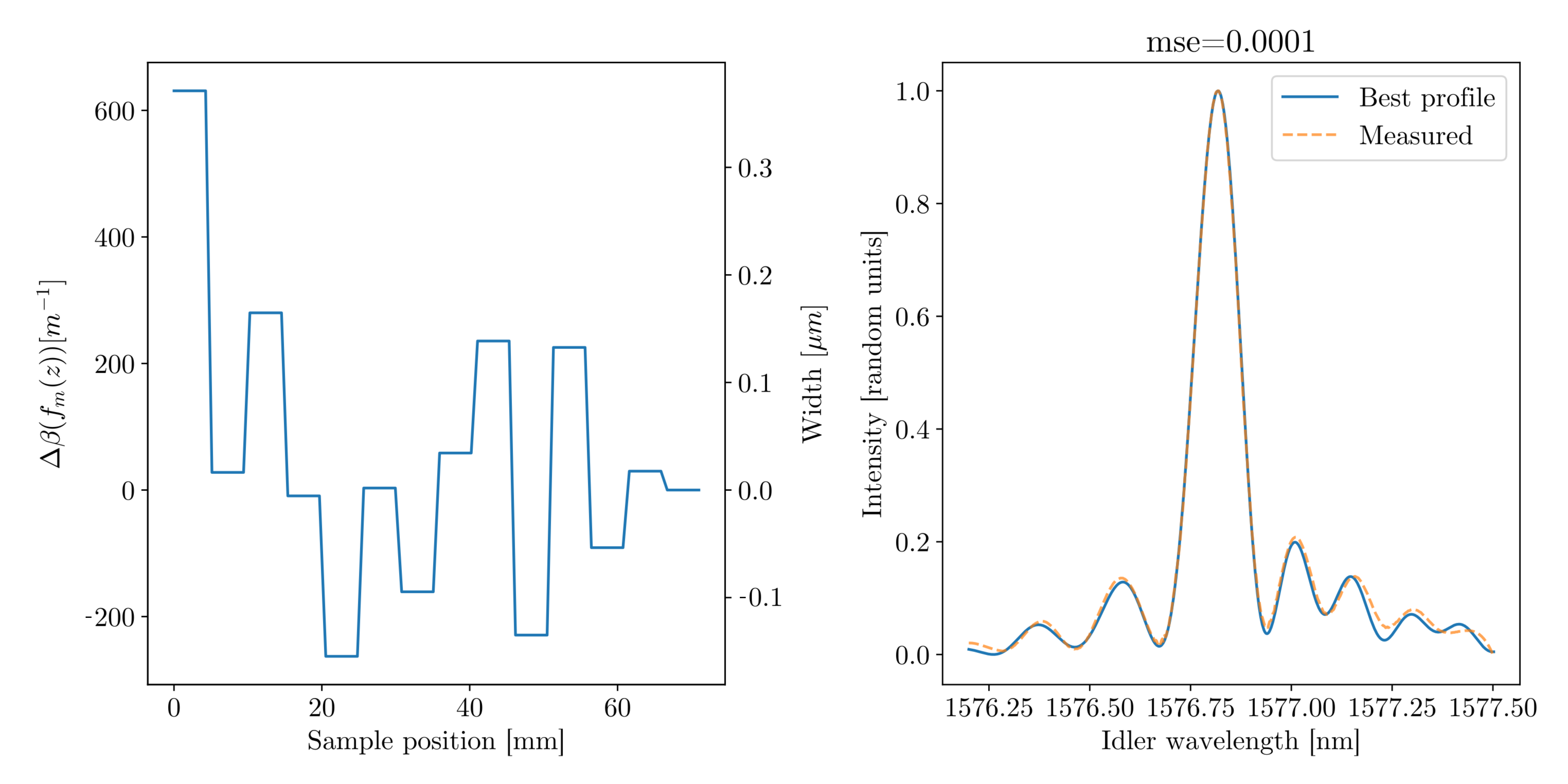}
	\caption{The resulting $\Delta\beta(f_m(z))$ profile after running the genetic algorithm is on the left. The y axis on the right are the equivalent waveguide width deviation for the profile. The best profile's phase-matching function is plotted against the measured one at room temperature on the right. The MSE is $0.0001$.}
	\label{fig:min_prof}
\end{figure}

We divide the waveguide in 14 sections of constant $f_m(z)$. 
This was chosen so that each section of the profile is smaller than the critical length for this process \cite{Santandrea2019}. 
This might cause oversampling of the problem and the solution not being unique but the overall profile is useful to quantify the magnitude of the inhomogeneities nonetheless.
We run our algorithm for a 100 generations with a population size of 100 profiles. 
The tournament has 4 participants. 
This took around 10 hours of computing time on 16 Gb of RAM. 
The best profile and resulting phase-matching spectrum can be seen in \refig{fig:min_prof}.

The algorithm produces a very good match with the measured phase-matching with a resulting MSE of 0.0001. 
This optimal profile is characterised by the presence of significant inhomogeneities, despite the measured spectrum being very close to the ideal one.
%On the phase-matching function of the resulting profile, we can see how there are significant inhomogeneities, although the bandwidth at room temperature was close to expected.  %although the experimental measurement was very good compared to the ideal phase-matching. 
This is a useful tool to estimate inhomogeneities in waveguides and gain a better knowledge of the fabrication process. 
The algorithm gave a similar result for another poled waveguide within the same sample.

\begin{figure}[h!]
	\centering
	\includegraphics[width=\linewidth]{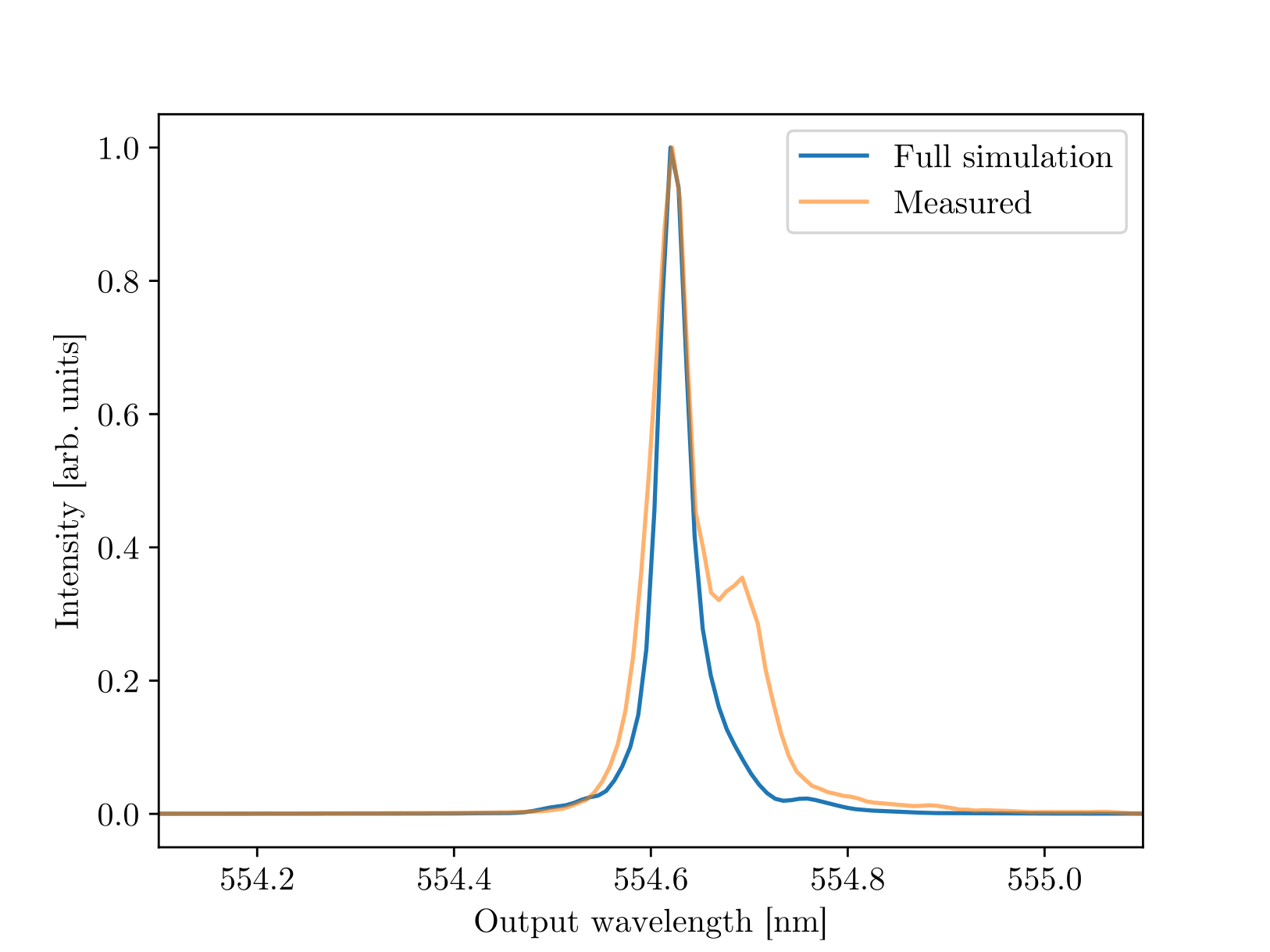}
	\caption{The simulated phase-matching function fits the main feature of the measured one.}
	\label{fig:full_sim}
\end{figure}

We can now combine the profile at room temperature and the effect of the high temperature 200 $^\circ$C. 
Henceforth, we simulate the full profile of the waveguide to estimate the experimental phase-matching function. 
The result is shown in \refig{fig:full_sim}.

The full simulation matches perfectly the main peak on the measured phase-matching function. 
The simulated bandwidth is very close to the measured one. 
The results are now fully explained by fabrication inhomogeneities and only the peak on the right side is unexpected. 
This may be caused by an inhomogeneity of the temperature distribution over the device on the copper oven.

%\textcolor{red}{We might need to ask Michael to help us sell this a bit better.} 
Overall, considering the bandwidth of the phase-matching spectrum at 200 degrees, this QPG device has a bandwidth compression of 16, doubling on previous results. 
The internal conversion efficiency of 20\% at \SI{4}{\milli\watt} pump power outperforms the current state of the art thanks to a fourfold increase in the sample length.
%The internal conversion efficiency was measured at 200 degrees by pumping the system with the maximum pump power available, \SI{4}{\milli\watt} and by measuring the depletion of the signal field through the device with the pump on and off. The investigated device shows a conversion efficiency of 20\%, above that of the device in \cite{Allgaier2017b} at the same pump power. 
Finally, the device is estimated to reach extinction ratios around \SI{21.5}{\deci\bel} by just filtering the asymmetry on the right side or addressing the temperature distribution.
Furthermore, if the asymmetries are addressed, a measurement with higher resolving power could reveal an ideal phase-matching function with \SI{35}{\deci\bel} extinction ratio or serve to further improve the fabrication process.

\section{Conclusions}

We analysed the current state of the art of non-linear wavelength converters using three benchmarks: conversion efficiency, bandwidth compression and extinction ratio.
This review reveals that current state-of-the-art samples do not reach the expected theoretical benchmark values due to fabrication errors.
%Although current devices have achieved good results, they are still far from expected values.

%The results of an improved device have been discussed.

To overcome current technological limit, a detailed analysis of the fabrication process has been performed. 
Consequently, the manufacturing process was drastically improved to correct sources of error in the fabrication of the devices and allow the fabrication of samples much larger than any comparable devices known to the authors.
The improved fabrication allowed us to produce a \SI{71}{\milli\meter} long QPG device, which has been shown to outperform previously published results at every benchmark under study.
%This produced a \SI{71}{\milli\meter} long QPG device that performs better that the current state of the art processes. 
This is due to a phase-matching spectrum very close to the ideal one and much narrower than any of the previous results thanks to the increased sample length. 
This immediately increases its bandwidth compression ratio to 16 and allows it to potentially reach the bandwidth of many current quantum memories. 
%Due to the much higher sample length, we have measured a fourfold increase in the internal conversion efficiency of our new device.
Due to the fourfold increase of the sample length, we have measured a conversion efficiency much higher than previous devices, even with a lower $\eta_{norm}$.
Finally, from the measured spectrum, we estimate \SI{21.5}{\deci\bel} extinction ratio. 
This pushes the discussed minimum estimable separation down to 0.5 $\sigma$ and allows interfacing with Raman quantum memories.
Moreover, having characterised a near-ideal phase-matching spectrum at room temperature, we expect the \SI{35}{\deci\bel} extinction ratio to be achievable by addressing the observed temperature inhomogeneities at higher temperatures.

A computational method to retrieve the $\Delta\beta$ profile of the non-linear process is given to infer the profile along the propagation direction of the devices. 
The profile can be obtained from the phase-matching function measurement, without any extra experimental effort. 
This profile can be use to compensate for the inhomogeneities found.

The study presented here highlights the importance of an optimized fabrication technology in order to achieve long and highly efficient QPG devices and that careful tuning of all the fabrication and experimental parameters involved can lead to a dramatic improvement of the desired non-linear process.
%Together with the method to easily retrieve the inhomogeneities on the device, they are a key to the production and characterisation of improved non-linear devices for quantum applications.

\section{Acknowledgments}

This work was funded by the Deutsche Forschungsgemeinschaft via SFB TRR 142.

\printbibliography
\end{document}